\documentclass[12pt]{article}

\usepackage{url}  % Formatting web addresses
\usepackage[latin1]{inputenc}
\usepackage{amsmath}
\usepackage{amsfonts}
\usepackage{graphicx}
\usepackage{float}

\newcommand{\Ber}{\text{Be}}
\renewcommand{\vec}[1]{\boldsymbol{#1}}
\newcommand{\prob}{\mathbb{P}}
\newcommand{\mat}[1]{\mathbf{#1}}
\newcommand{\expect}{\mathbb{E}}

\begin{document}

\title{ProbCD: enrichment analysis accounting for categorization uncertainty}

\author{Ricardo Z.N. Vêncio$^{1}$\footnote{to whom correspondence should be addressed: rvencio@gmail.com}, Ilya
Shmulevich$^1$} \footnotetext[1]{Institute for Systems Biology, 1441 North 34th street, Seattle, WA 98103-8904, USA}

\maketitle

\begin{abstract}

As in many other areas of science, systems biology makes extensive use of statistical association and significance estimates in contingency tables, a
type of categorical data analysis known in this field as enrichment (also over-representation or enhancement) analysis. In spite of efforts to create
probabilistic annotations, especially in the Gene Ontology context, or to deal with uncertainty in high throughput-based datasets, current enrichment
methods largely ignore this probabilistic information since they are mainly based on variants of the Fisher Exact Test.  We developed an open-source
R package to deal with probabilistic categorical data analysis, ProbCD, that does not require a static contingency table. The contingency table for
the enrichment problem is built using the expectation of a Bernoulli Scheme stochastic process given the categorization probabilities. An on-line
interface was created to allow usage by non-programmers and is available at: http://xerad.systemsbiology.net/ProbCD/. We present an analysis
framework and software tools to address the issue of uncertainty in categorical data analysis. In particular, concerning the enrichment analysis,
ProbCD can accommodate: (i) the stochastic nature of the high-throughput experimental techniques and (ii) probabilistic gene annotation.

\end{abstract}

\section*{Background}

The system-level approach to data analysis known as enrichment analysis (also known as over-representation or enhancement analysis) is now
commonplace. Moreover, the number of available software tools to perform such analysis is large (see \cite{dopazo2006fim,Rivals2007} for
comprehensive reviews). The preferred way to formalize the enrichment problem is by means of a contingency table, often $2 \times 2$.

~

The mathematical problem is conceptually generic, being applied to diverse types of data, such as genomics, transcriptomis or proteomics datasets;
diverse types of analysis, including multiple and/or ordered outcomes; and diverse types of gene classification schemes, such as Gene Ontology (GO),
KEGG or organism-specific ones. For a given ontology term $t$ defining the set of genes $G_t$ and its complementary set $G_t^c$, the general
enrichment analysis contingency table is:

~

\begin{tabular}{c|cc}
 &  $G_t$    &   $G_t^c$   \\
\hline
$outcome_1$   & $X_{1,1}$ & $X_{1,2}$ \\
$outcome_2$   & $X_{2,1}$ & $X_{2,2}$ \\
$\cdots$      & $\cdots$  & $\cdots$  \\
$outcome_k$   & $X_{k,1}$ & $X_{k,2}$ \\
\end{tabular}

~

Besides measuring the statistical significance of the null hypothesis that the rows and columns are independent, as yielded by Fisher's Exact Test
\cite{fisher1922ichic} and Fisher-like methods \cite{dopazo2006fim,Rivals2007}, it is also possible to measure statistical association between a
table's rows and columns \cite{goodman1954mac} (a detailed discussion on significance vs. association in the enrichment problem context can be found
in \cite{vencio2006bba}).

~

Most of the attention in the enrichment analysis problem has focused on issues such as the search for the best multiple-test correction or the
implementation of better user-friendly software interfaces to facilitate biologist's exploratory work \cite{dopazo2006fim}. However, one of the
limitations that the available approaches still share is that they assume, explicitly or implicitly, that one is able to construct the contingency
table exactly, without uncertainty in populating its cells.

~

Recently, the computational biology community has been witnessing an increasing interest in probabilistic approaches to gene annotation, particularly
in the Gene Ontology (GO) context, as a realization of the limitations imposed by the traditional deterministic and context-independent gene
annotation schemes \cite{joshi2004gsg, levy2005pap, engelhardt2005pmf, martin2005gnm, engelhardt2006gmp, carroll2006pcu, vinayagam2006gta,
jones2007eae}. These efforts are motivated by: the necessity to assess the error propagation in automatic gene annotation \cite{levy2005pap,
jones2007eae}; desire to include different types of evidence sources such as protein-protein interaction \cite{joshi2004gsg, carroll2006pcu} or
phylogenomics \cite{engelhardt2005pmf, engelhardt2006gmp} and annotation extrapolation from model organisms to others \cite{vinayagam2006gta,
martin2005gnm}. Meanwhile, the probabilistic nature of data obtained by high-throughput measurement techniques is well recognized and a number of
attempts to model it were proposed over the past decade in various experimental contexts \cite{book1,book2}.  However, these efforts are not
integrally taken into account when usual enrichment analysis is performed.

~

We describe a computational solution that is able to deal with the uncertainty introduced in enrichment analysis due to: (i) the
stochastic nature of the results obtained with such high-throughput experimental techniques or (ii) probabilistic gene annotation.

%%%%%%%%%%%%%%%%%%%%%%
\section*{Implementation}

ProbCD is an open-source software designed to perform probabilistic categorical data analysis. ProbCD is
written in R \cite{Rwebsite} with a level of modularity that makes it suitable to be incorporated by existing
development efforts of integrative tools \cite{shannon2006gos}. To facilitate the usage by researchers with no
knowledge of R, we implemented a user-friendly web-based interface for the software, which is not
limited to any particular organism. The on-line interface and the source-code are available on the project's
website \cite{baygo2web}.

~

The idea behind ProbCD's implementation is to formally represent the intuitive process of building a contingency table in a probabilistic manner.
Informally speaking, each element to be placed in the contingency table is not considered to be indivisible, but instead is ``shared", according to
probabilistic rules, among the contingency table's cells in a manner that is conceptually similar to fuzzy membership. The theoretical and
computational implementation aspects are described in detail below.

~

Without loss of generality, the following descriptions are applied considering one particular ontology term $t$ that is associated with a set of
genes, named simply as $G_t$. It should be noted that $G_t$ is not restricted to the Gene Ontology categorization and can be any kind of
classification or annotation.

~

The vector $\vec q$ contains a probabilistic annotation for all $g$ of the organism's genes: $q_j = \prob(gene_j \in G_t)$ for $j \in \{1, \cdots,
g\}$.  This probabilistic annotation is assumed to be given, typically obtained from some analysis process. The deterministic scenario corresponds
simply to $\prob(gene_j \in G_t) \in \{0,1\}$, and hence is a special case.

~

The matrix $\mat P$ contains a probabilistic description for all $k$ possible outcomes of the property being studied. Therefore, $\mat P$ is a $k
\times g$ matrix with elements $P_{i,j} = \prob(gene_j \in outcome_i)$ for $j \in \{1, \cdots, g\}$ and $i \in \{1, \cdots, k\}$. This probabilistic
description of the data uncertainty is assumed to be given.

~

To motivate the general probabilistic model, it is useful to examine an arbitrary $2 \times 2$ example in the deterministic scenario:

~

\begin{tabular}{c|cc}
 &  $G$    &   $G^c$   \\
\hline
$H$   & $x_{1,1}$ & $x_{1,2}$ \\
$H^c$   & $x_{2,1}$ & $x_{2,2}$ \\
\end{tabular}

~

where all $x$'s are the counts of a regular contingency table over the gene sets $G$ and $H$. In its matrix representation:
\[
\left( \begin{array}{cc}
x_{1,1} & x_{1,2} \\
x_{2,1} & x_{2,2} \\
\end{array} \right)
=
\left( \begin{array}{cc}
\sum_{j} \vec 1_{\{gene_j \in H\}} \vec 1_{\{gene_j \in G\}} & \sum_{j} \vec 1_{\{gene_j \in H\}} \vec 1_{\{gene_j \in G^c\}}   \\
\sum_{j} \vec 1_{\{gene_j \in H^c\}} \vec 1_{\{gene_j \in G\}}  & \sum_{j} \vec 1_{\{gene_j \in H^c\}} \vec 1_{\{gene_j \in G^c\}}   \\
\end{array} \right)
\]
where $\vec 1_{\{\}}$ is the indicator function.

~

Inspired by this representation, it is easy to see that the ``hard" indicator functions may be substituted by Bernoulli random variables in order to
account for the categorization uncertainty.  Since all sets are finite, the indicator functions can be represented as vectors in $\{0,1\}^g$ and the
sums over all genes as dot products. In a generic scenario, with given non-deterministic $\mat P$ and $\vec q$, the contingency table represented by
$\mat X| \mat P, \vec q$ is a random matrix that is difficult to describe in closed form. It is also not compatible with the statistical formalism
supporting Fisher's Exact Test or other well-known Fisher-like approaches, as these are not applicable to random tables. 

~

The contingency table is defined in terms of Bernoulli Schemes \cite{bernuolliwiki} which is the generalization of the Bernoulli Process to more than
two possible outcomes. The notation $\vec Z \sim \Ber(p_1, \cdots, p_n)$ represents the distribution:
\[ \vec z = \left\{ \begin{array}{ll}
        (1,0,0, \cdots, 0) & \mbox{with probability $p_1$};\\
        (0,1,0, \cdots, 0) & \mbox{with probability $p_2$};\\
        (0,0,1, \cdots, 0) & \mbox{with probability $p_3$};\\
        \cdots \\
        (0,0,0, \cdots, 1) & \mbox{with probability $p_n$}.\\ p_1 + \cdots + p_n = 1\end{array} \right. \]

The random variable $\mat X$ is a matrix representation of a $k \times 2$ contingency table:
\[
\left( \begin{array}{cc}
X_{1,1} & X_{1,2} \\
\cdots & \cdots \\
X_{k,1} & X_{k,2} \\
\end{array} \right)
=
\left( \begin{array}{cc}
\vec d_1 \cdot \vec a_1  & \vec d_1 \cdot \vec a_2 \\
\cdots & \cdots \\
\vec d_k \cdot \vec a_1  & \vec d_k \cdot \vec a_2 \\
\end{array} \right)
\]
where $\cdot$ is the usual dot-product, 
$\vec a_i = (A_{i,1}, \cdots, A_{i,g})$ is a row-vector of a $2 \times g$ binary matrix $\mat A$ such that $(A_{1,j},A_{2,j})|q_j \sim 
\Ber(q_j,1-q_j)$ and $\vec d_i = (D_{i,1}, \cdots, D_{i,g})$ is a row-vector of a $k \times g$ binary matrix $\mat D$ such that $(D_{1,j}, 
\cdots, D_{k,j})|(P_{1,j}, \cdots, P_{k,j}) \sim \Ber(P_{1,j}, \cdots, P_{k,j})$.

~

It is very easy to extend this framework for completely 
generic $k \times m$ tables ($m > 2$), but this would be outside the scope of the ontology enrichment problem.

~

To measure statistical association between rows and columns in an ordered contingency tables, as is analogously made when correlations are calculated
for non-categorical data, Goodman-Kruskal's gamma $\gamma(\mat X)$ can be used \cite{goodman1954mac,garson1976psm,vencio2006bba}. ProbCD calculates
the statistical association accounting for the stochastic nature of the table's categorization reporting $\gamma = \gamma(\expect[\mat X| \mat P,
\vec q])$, where $\expect$ is the expectation operator. This is a particular association measurement that can be easily changed for other
user-implemented options.

~

The dichotomous case, which is the simplest one, gives a more intuitive illustration on how the association is calculated in practice for the
particular implementation: $\expect[X_{1,1}| \mat P, \vec q] = E_{1,1} = P_{1,1} q_1 + \cdots + P_{1,g} q_g$, $\expect[X_{2,1}| \mat P, \vec
q] = E_{2,1} = (1-P_{1,1}) q_1 + \cdots + (1-P_{1,g}) q_g$, $\expect[X_{1,2}| \mat P, \vec q] = E_{1,2} = P_{1,1} (1-q_1) + \cdots + P_{1,g}
(1-q_g)$, $\expect[X_{2,2}| \mat P, \vec q] = E_{2,2} = (1-P_{1,1}) (1-q_1) + \cdots + (1-P_{1,g}) (1-q_g)$ and $\gamma = (E_{1,2}E_{2,2} -
E_{1,2}E_{2,1})/(E_{1,2}E_{2,2} + E_{1,2}E_{2,1})$.

~

To measure the statistical significance of the estimated association, ProbCD uses a randomization approach. The null distribution for the
Goodman-Kruskal's gamma, $\gamma^*$, is proposed to be estimated from several permutation rounds. In each round a gene $j$ receives randomly its
probabilities $(P_{1,j}^*, \cdots, P_{k,j}^*)$ from one of the $g$ possible columns of $\mat P$ and an association value is calculated. The
significance of the statistical association between rows and columns in the contingency table is calculated as $p = \prob(\gamma^* \geq \gamma)$. A
term $t$ is significantly over-represented (or equivalently, the gene list is enriched for $t$) depending on user-defined thresholds
for significance and/or association.

%%%%%%%%%%%%%%%%%%%%%%
\section*{Results}

The following examples illustrate the potential utility of considering probabilistic annotations and/or data uncertainty assessment in the
enrichment analysis using ProbCD on artificial datasets and a published yeast dataset.

~

The point of the following illustration is to show that even ontology terms annotated with modest probabilities can be considered to be
over-represented if the list of genes obtained behave in a supportive pattern. Consider a hypothetical organism with
100 genes annotated in several GO terms, as described in the Additional Files. The genes $gene_1$ to $gene_{20}$ are deterministically
annotated to the ontology term $t = a$. In other words, assume that it is well known that these 20 genes have some given functionality $a$. 
The
experiment, for example from a hypothetical proteomics dataset, yielded a deterministic list of differentially expressed (DE) genes ranging 
from
$gene_1$ to $gene_{10}$. The contingency table for this problem is, therefore:

~

\begin{tabular}{c|cc}
 &  $G_a$    &   $G_a^c$   \\
\hline
$DE$   & 10 & 0 \\
$DE^c$   & 10 & 80 \\
\end{tabular}

~

In this case, the gene list is clearly enriched for $a$ within any meaningful significance cutoff. Consider now a second ontology term $b$ obtained
from a probability-based source with $\prob(gene_i \in G_b) = 40\%$, $i \in \{1,\cdots,20\}$. A probability of only 40\% generally would not be
sufficient evidence to warrant the inclusion of those 20 genes in $G_b$ considering a usual deterministic framework and, therefore, would not be
analyzed by deterministic-based methods, such as the Fisher's Exact Test. However, ProbCD is able to incorporate this information and yields: $\gamma
= 0.87$ and $p < 10^{-4}$ in 10000 permutation rounds, a significant enrichment for $b$. One can easily imagine, for example, genes that have a main
function $a$ but also have a different function $b$ in, say, $40\%$ of documented conditions.

~

The point of the following illustration is to show that the incorporation of probabilistic annotation information does not always translate to
addition of terms into the enrichment result, as in the example above, but it can also mean the exclusion of non-relevant terms. Consider a
hypothetical organism with 1100 genes. Let the genes $gene_1$ to $gene_{100}$ be grouped together in a cluster $H$ after some genomic sequence
analysis. Let the term $a$ be annotated deterministically (Additional Files) yielding the contingency table:

~

\begin{tabular}{c|cc}
 &  $G_a$    &   $G_a^c$   \\
\hline
$H$   & 100 & 0 \\
$H^c$   & 100 & 900 \\
\end{tabular}

~

In this situation, $H$ is clearly enriched for $a$ within any meaningful significance cutoff. Let now the same annotation incorporate some evidence
levels by defining: $\prob(gene_i \in G_a) = 99\%$ for $i \in \{1,\cdots,10\}$ and $\prob(gene_i \in G_a) = 1\%$ for $i \in \{11,\cdots,100\}$.
Intuitively, this means that only 10 out of 100 genes clustered in $H$ are, in fact, confidently annotated with the ontology term $a$. The
incorporation of this information results in non-significant enrichment of $H$ for $a$ since: $\gamma = 0.0425$ and $p = 0.42$ in 1000 permutation
rounds.  Therefore, it can be useful to incorporate uncertainty information into the enrichment analysis to also down-rank potentially spurious
enrichment results.

~

The purpose of the following illustration is to show the impact of considering the uncertainty in lists of genes, rather than in the annotations, on
the enrichment analysis. In this example, the aim is to find which GO terms, annotating the yeast \emph{Saccharomyces cerevisiae}, are statistically
associated with periodic expression levels, measured by microarray technology \cite{andersson2006bdp}. Andersson and colleagues
\cite{andersson2006bdp} devised a Bayesian model that produces the probability that a gene is periodically expressed during the cell-cycle. This
simple presentation is sufficient for our objectives in this work, but the interested reader can find more details (e.g. the definition of
``periodic", etc.) in the original work. In this example, the annotation is considered to be deterministic and it was downloaded from the GO project 
page (March 2007) \cite{goyeast}.

~

To perform the usual enrichment analysis one needs to define a probability cutoff value in order to split the gene list in two: the periodic genes
and the non-periodic genes. Consider initially the reasonable cutoff $\prob(gene_i$ is periodic$) \geq 70\%$ and focus on a single GO term
GO:0007090 (regulation of S phase of mitotic cell cycle), defined as \emph{``a cell cycle process that modulates the frequency, rate or
extent of the progression through the S phase of mitotic cell cycle"}. Although this GO term is clearly associated with periodic gene expression,
performing a usual enrichment analysis results in the conclusion that the periodic genes are not significantly enriched for GO:0007090 within usual
significance cutoffs ($p$-value = 0.065). 

~

\begin{figure}[H]
\center
\includegraphics[scale=0.5]{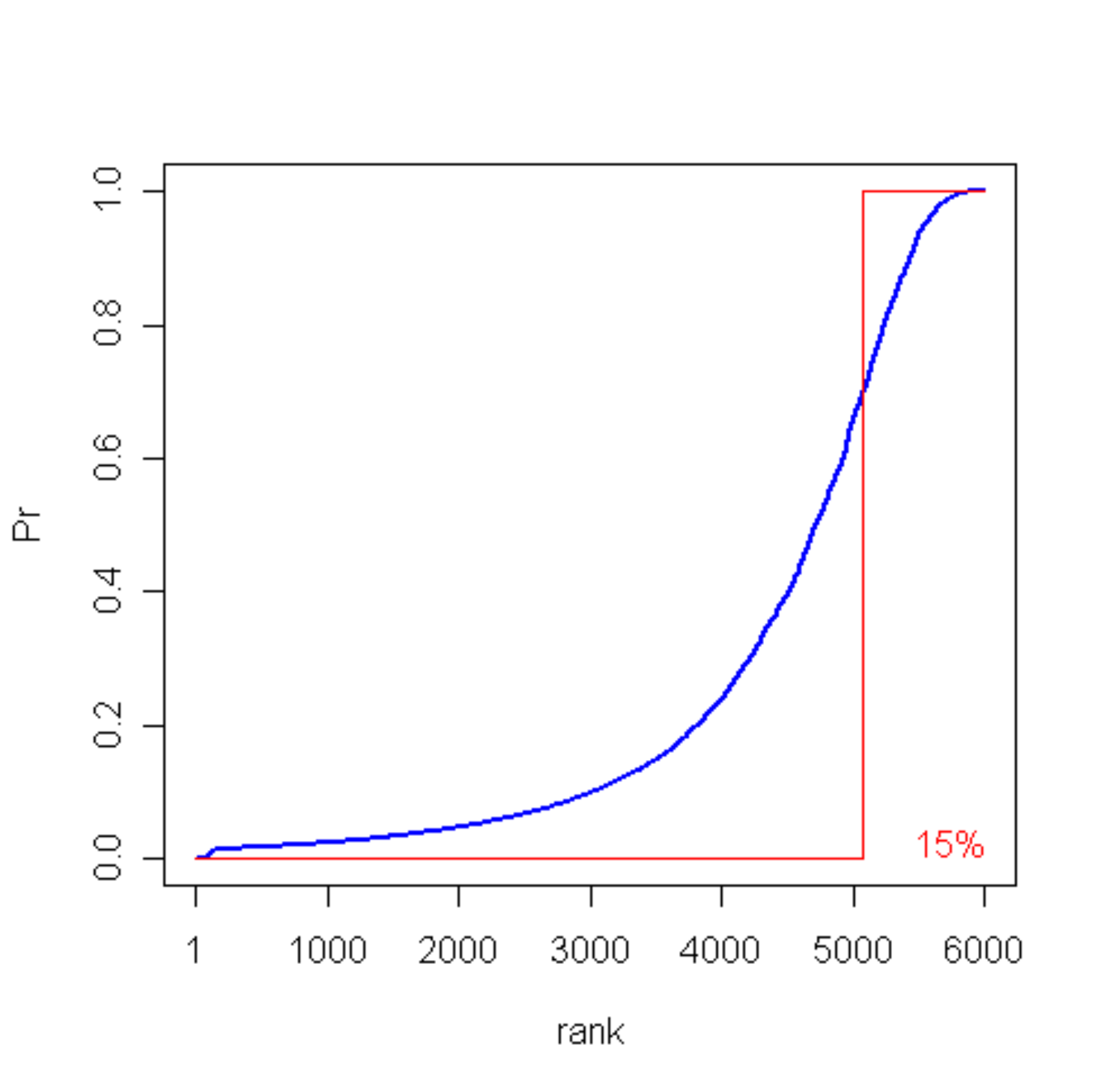}
\caption{
Probability of being periodic. The blue curve represents the probability of a gene being periodic (Pr) according to the model of
\cite{andersson2006bdp}. The genes are sorted by probability values (rank) on the horizontal axis to facilitate the visualization. The red curve is
the deterministic approximation using a 70\% probability cutoff to consider a gene as periodic: $\prob(gene_i$ is periodic$) \geq 0.70 \Rightarrow
\prob(gene_i$ is periodic$) = 1$ and $\prob(gene_i$ is periodic$) < 0.70 \Rightarrow \prob(gene_i$ is periodic$) = 0$. This approximation labels 15\%
of the genes as periodic.
}\label{fig:corte70porcento}
\end{figure}

Suspecting that this non-intuitive result could be due to the probability threshold chosen to select periodic
genes, illustrated in the Figure \ref{fig:corte70porcento}, one could repeat the same analysis above building
the contingency table considering the cutoffs $\prob(gene_i$ is periodic$) \geq 50\%$, 95\%, 99\% or 99.99\%.
The result of this repeated analysis is also non-intuitive since the $p$-values are: 0.12, 1.0, 1.0 and 1.0 for
50\%, 95\%, 99\% and 99.99\% cutoffs, respectively, meaning that increasing the stringency to define a gene as
periodic only decreases the significance of the enrichment for GO:0007090.

~

Using ProbCD one can consider the actual probability of being periodic (blue curve in Figure
\ref{fig:corte70porcento}) in the enrichment analysis instead of using the deterministic approximation (red curve
in Figure \ref{fig:corte70porcento}). This result in a relatively high statistical association between
periodicity and the term ``regulation of S phase of mitotic cell cycle" ($\gamma$ = 0.78) also with high
significance ($p$ = 0.009 in 1000 simulation rounds). Judging subjectively by the definition of GO:0007090,
ProbCD returned a meaningful result.

~

Other similar cases can be easily identified. For example, the GO term GO:0000083 (G1/S-specific transcription in mitotic cell cycle) exhibits
erratic behavior depending on the chosen cutoff for the probability of being periodic: $p$-value of 0.15, 0.10, 0.01, 0.096 and 1.0 for 50\%, 75\%,
95\%, 99\% and 99.99\% cutoffs, respectively. The probability stringency used to build the contingency table and the subsequent significance test are
not necessarily correlated. ProbCD yielded a significant ($p$ = 0.006) moderate association ($\gamma$ = 0.48) for GO:0000083. Other examples include
GO:0045787 (positive regulation of progression through cell cycle), defined as \emph{``any process that activates or increases the frequency, rate or
extent of progression through the cell cycle"}, which would be called significant using the regular enrichment method only if the right probability
cutoff $\prob(gene_i$ is periodic$) \geq 95\%$ is guessed initially: $p$-value of 0.047, 0.024, 0.0058, 0.086 and 0.024 for 50\%, 75\%, 95\%, 99\%
and 99.99\%, respectively.

~

\begin{figure}[H]
\center
\includegraphics[scale=0.5]{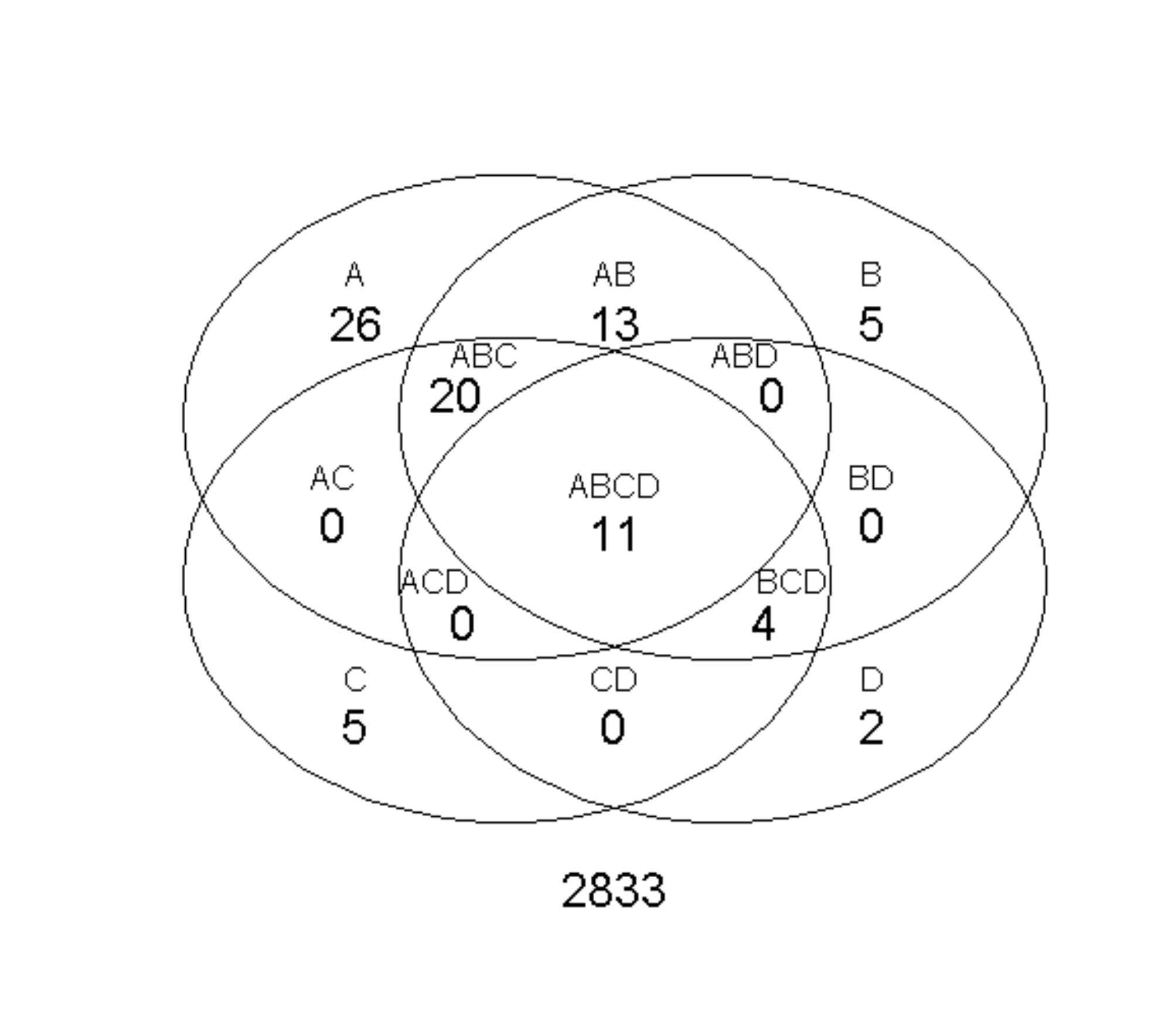}
\caption{Venn diagram of over-represented terms. 
The Venn diagram shows the number of GO terms considered significantly over-represented ($p$-value $\leq$ 0.01) by the Fisher Exact Test using four
different probability cutoffs $\prob(gene_i$ is periodic$) \geq$ A, B, C or D $\Rightarrow$ periodic: A = 0.70, B = 0.95, C = 0.99 and D =
0.9999.
}\label{fig:venn4}
\end{figure}

The above analysis process is repeated for all GO terms, with the results available as Additional Files and summarized in Figure
\ref{fig:venn4}.

~

This figure suggests that there is a large variability in the possible final outcome of an enrichment analysis depending on the probability cutoff
used to build the associated contingency table. This variability is avoided by ProbCD because it directly takes into account the uncertainty in the 
data instead of introducing a discretization step (Figure \ref{fig:corte70porcento}).

~

\begin{figure}[H]
\center
\includegraphics[scale=0.5]{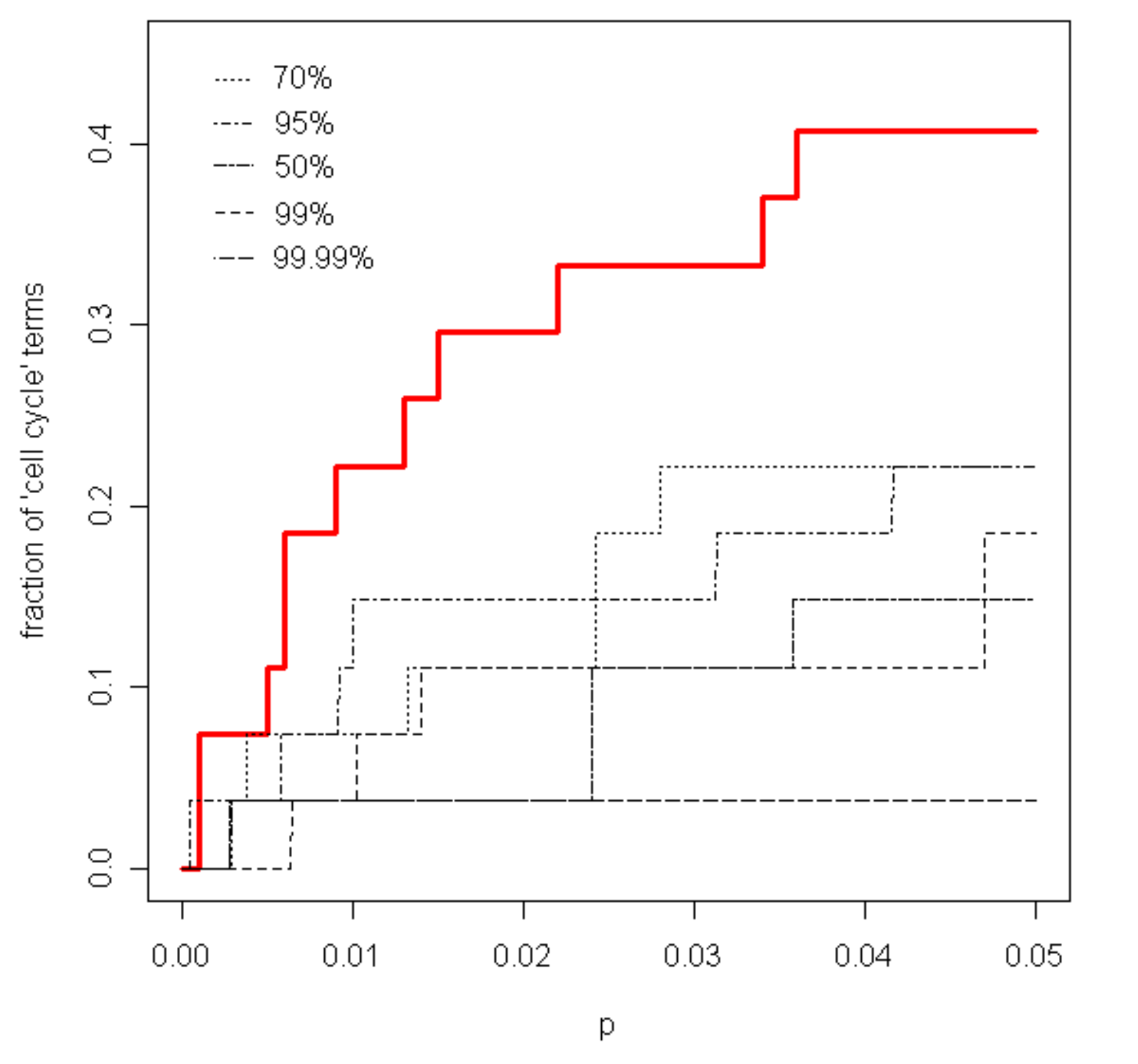}
\caption{
Fraction of ``cell-cycle" GO terms selected as a function of the $p$-value.
The curves show the fraction of GO terms containing the word ``cell-cycle" in their definition that are considered
significant as a function of the significance cutoff ($p$). The red curve is obtained with ProbCD and all others
are obtained with one of the probability cutoffs: 50\%, 70\%, 95\%, 99\% or 99.99\%.
}\label{fig:roclike}
\end{figure}

Figure \ref{fig:roclike} shows that ProbCD considers more terms (vertical axis in Figure \ref{fig:roclike}) containing the word ``cell cycle",
likely associated to periodically expressed genes, as significant if compared to the usual enrichment analysis in a wide range of significance values
($p$ in Figure \ref{fig:roclike}). Although this is not a proof, since one cannot be certain about which ``cell cycle"-marked terms
should be enriched, we argue that this is a reasonable indication that one can, in fact, avoid the discretization step when building the enrichment
problem using ProbCD and obtain meaningful results.

~

%%%%%%%%%%%%%%%%%%%%%%
\section*{Discussion and Conclusions}

The usual enrichment analysis is a particular case in this probabilistic framework and can be obtained by ProbCD ignoring the difference between
evidence sources in gene annotation and defining fixed gene lists, which would correspond to the deterministic setting: $q_j = \prob(gene_j \in G_t)
= 1$ or $0$ and $P_{i,j} = \prob(gene_j \in outcome_i) = 1$ or $0$. 

~

Even if a probabilistic annotation is not readily available for a given organism, it could be interesting to perform enrichment analysis taking into
account some form of weighting on available annotations according to their reliability. For a concrete example, the GO Consortium \cite{GOpage}
provides annotations accompanied with evidence codes related to the kind/level of evidence available for a given GO annotation \cite{GOevidence},
such as \emph{IEA: Inferred from Electronic Annotation}, \emph{IMP: Inferred from Mutant Phenotype}, \emph{RCA: inferred from Reviewed Computational
Analysis} or \emph{IDA: Inferred from Direct Assay}. It is known that some evidence sources are more reliable than others and this knowledge can be
used, in a Bayesian sense, as subjective probabilities.

~

Once an annotation is considered in a probabilistic framework, it could reflect a dependence on the context. One can consider cases in which
$\prob(gene_j \in G_t|$ disease $) \gg \prob(gene_j \in G_t)$, defining context-dependent gene annotations derived, for instance, from automatic
literature mining \cite{aubry2006ceb}. 

~

Our intention is to complement existing approaches, rather than substitute them. Toward this aim, we built ProbCD to be as modular as possible in
order to be incorporated into existent software or pipelines \cite{shannon2006gos}, composed of ontology pre-processing
\cite{lewin2006ggo} or powerful visualization capabilities \cite{maere2005bcp, sealfon2006gig}. 

~

It is important to note that ProbCD is also applicable to other categorical data analysis contexts in which the construction of contingency tables is
subject to uncertainty, a recurrent theme in science.

%%%%%%%%%%%%%%%%%%%%%%%%%%%%%%%%
\section*{Authors contributions}

RZNV implemented the project. IS supervised the project. All authors read and approved the final manuscript.

%%%%%%%%%%%%%%%%%%%%%%%%%%%
\section*{Acknowledgments}

We thank Drs. John Boyle, Vesteinn Thorsson, Nathan Price and other ISB colleagues for insightful discussions. This work is partially supported by
NIH grants U54-AI54253, U19-AI057266 and P50-GMO-76547.

%%%%%%%%%%%

\bibliography{bmc_article}

%%%%%%%%%%%

\end{document}